\documentclass{an}
\usepackage{graphicx}
\usepackage{times}
\usepackage{fancyhdr}
\newcommand{\st}[2]{\stackrel{\mbox{\tiny (#1)}}{#2}\hspace{-0.1cm}}

\begin{document}

\title{Magnetic field generation from cosmological perturbations}

\author{Keitaro Takahashi$^{1}$, Kiyotomo Ichiki$^{2}$,
Hiroshi Ohno$^{3}$, Hidekazu Hanayama$^{2,4}$ and Naoshi Sugiyama$^{2}$}
\institute{
Department of Physics, Princeton University, Princeton,
NJ 08544.
\and
National Astronomical Observatory of Japan, Mitaka,
Tokyo 181-8588, Japan.
\and
Laboratory, Corporate Research and Development Center,
Toshiba Corporation, 1, Komukai Toshiba-cho, Saiwai-ku, Kawasaki 212-8582, Japan.
\and
Department of Astronomy, School of Science,
University of Tokyo, Hongo 7-3-1, Bunkyo, Tokyo 113-0033, Japan.
}

\date{Received; accepted; published online}

\abstract{
We discuss generation of magnetic field from cosmological perturbations.
We consider the evolution of three component plasma (electron, proton and photon)
evaluating the collision term between elecrons and photons up to the second order.
The collision term is shown to induce electric current, which then generate magnetic field.
There are three contributions, two of which can be evaluated from the first-order quantities,
while the other one is fluid vorticity which is purely second order. We compute numerically
the magnitudes of the former contributions and shows that the amplitude of the produced
magnetic field is about $\sim 10^{-19} {\rm G}$ at 10kpc comoving scale at present.
Compared to astrophysical and inflationary mechanisms for seed-field generation,
our study suffers from much less ambiguities concerning unknown physics and/or processes.
}

\correspondence{email:ktakahas@Princeton.EDU}

\maketitle

\section{Introduction}

There are convincing evidences that imply existence of substantial magnetic fields
in various astronomical objects. Not only galaxies but systems with even larger scales,
such as cluster of galaxies and extra-cluster fields, have their own magnetic fields
(for a review on cosmological magnetic fields, see e.g., [\cite{Widrow02}]).
Conventionally the magnetic fields in galaxies, and possibly in clusters of galaxies,
are considered to have been amplified and maintained by dynamo mechanism.
However, the dynamo mechanism needs the seed magnetic field and does not explain the origin
of the magnetic fields.

There have been many attempts to generate the seed fields. One of the approaches
to this problem is to generate magnetic fields astrophysically, often involving
the Biermann mechanism [\cite{Biermann50}]. This mechanism has been applied to various
systems: large-scale structure formation [\cite{Kulsrud97}], ionizing front [\cite{Gnedin00}],
protogalaxies [\cite{Davies00}] and supernova remnant of the first stars [\cite{Hanayama05}].
These studies showed the possibilities of magnetic-field generation with amplitudes of order
$10^{-16} \sim 10^{-21} {\rm G}$, which would be enough for the required field
$\sim 10^{-20} {\rm G}$. 

On the other hand, cosmological origins, which are often concerned with inflation,
can produce magnetic field with coherence lengths of much larger scales, which is typically
the horizon scale or possibly super-horizon scale
[\cite{Turner88,Bamba04,Ashoorioon04,Bertolami99,Bertolami05}]. Also they can often produce
fields with a wide range of length scales, while astrophysical mechanisms can produce fields
only with their characteristic scales. For a constraints on magnetic field with
cosmological scale, see [\cite{Yamazaki04}].

Here, we consider magnetic-field generation from cosmological perturbations
after inflation. Around and after the decoupling, coupling among charged particles
and photons become so weak that electric current can be induced by the difference
in motions of protons and electrons. This electric current leads to generation of
magnetic fields. It is well known that vorticity of plasma produces such electric current
[\cite{Harrison70,Hogan00,Berezhiani04,Matarrese04,Gopal04,Lesch95}]. However, because vorticity
is not produced at the first order in cosmological perturbations, we must study
the second order. We will consider equations of motion for protons, electrons and
photons separately up to the second order, although equation of motion for photons
does not appear explicitly. To study electric current appropriately, we must treat
the three components separately [\cite{Gopal04}]. Furthermore, we will
evaluate the collision term between electrons and photons, which is dominant and
essential for the magnetic-field generation, up to the second order. From the equations
of motion for protons and electrons, we obtain a generalized Ohm's law, which, combined
with the Maxwell equations, leads to an evolution equation for magnetic field.
Our study is based on the cosmological perturbation theory, which is highly successful
in the anisotropies of the cosmic microwave background, and suffers from much less ambiguities
concerning unknown physics and/or processes compared to astrophysical and inflationary
mechanisms for seed-field generation. For details of our study, see [\cite{KT05,KT06}].

\begin{figure}
\resizebox{\hsize}{!}
{\includegraphics[width=1cm,clip]{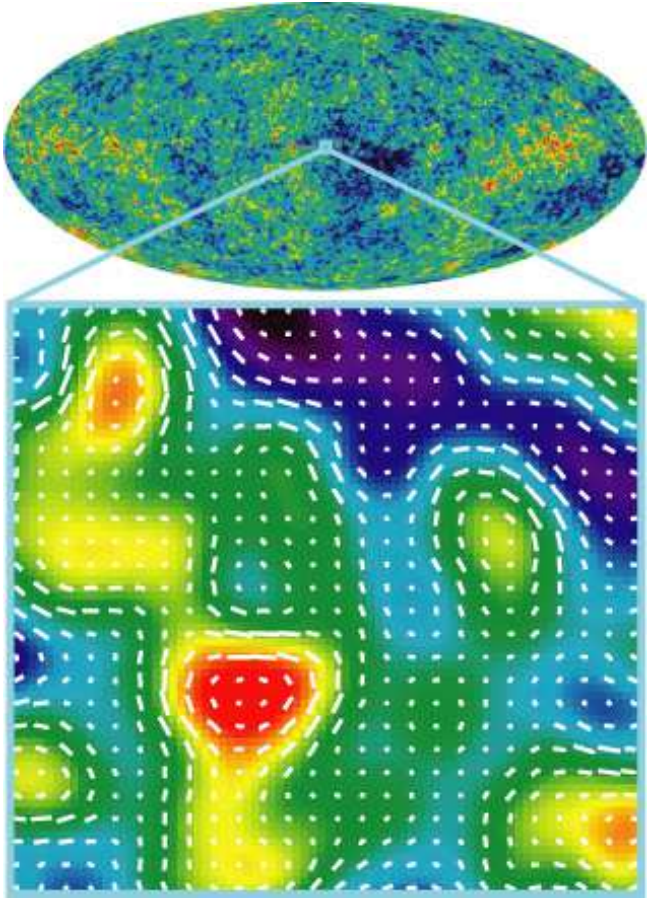}}
\caption{All sky map of cosmological microwave background
anisotropy obtained by the Wilkinson Microwave Anisotropy Probe (WMAP)
satellite (upper) and schematic picture of cosmological magnetic fields generated
from density fluctuations discussed here.
Red  (blue) regions  are hot (cold) spots with a range 
of temperature $\sim 2.725$ K $\pm 200 \mu $ K.
The magnetic field vectors are shown together with the map. 
Strong magnetic fields are generated by currents where
the gradient of density perturbation in photons is large.
\label{fig:image}}
\end{figure}

\section{Formulation}

Euler equations for proton fluid and electron fluid are given by
\begin{eqnarray}
&& ( \delta^{i}_{~\mu} + u^{i} u_{\mu} )
   \left( T^{\mu\nu}_{p ~ ;\nu} + T^{\mu\nu}_{{\rm EM}p ;\nu} \right)
   = C^{(C)i}_{pe} + C^{(T)i}_{p \gamma},
\label{eq:EOMp} \\
&& ( \delta^{i}_{~\mu} + u^{i} u_{\mu} )
   \left( T^{\mu\nu}_{e ~ ;\nu} + T^{\mu\nu}_{{\rm EM}e ;\nu} \right)
   = C^{(C)i}_{ep} + C^{(T)i}_{e \gamma},
\label{eq:EOMe}
\end{eqnarray}
where $T^{\mu\nu}_{p(e)}$ and $T^{\mu\nu}_{{\rm EM}p(e)}$ are the energy-momentum
tensor of proton (electron) fluid and electromagnetic field coupling to protons (electrons)
current, respectively. Here $\mu, \nu = 0,1,2,3$ and $i = 1,2,3$.
The projection of the divergence of the energy-momentum tensors are computed as
\begin{eqnarray}
&& ( \delta^{i}_{~\mu} + u^{i} u_{\mu} ) T^{\mu\nu}_{~~;\nu} \nonumber \\
&& ~~~~
   = ( \rho + p ) u^{\mu} u^{i}_{~;\mu}
     + ( \delta^{i}_{~\mu} + u^{i} u_{\mu} ) P^{,\mu},
\\
&& ( \delta^{i}_{~\mu} + u^{i} u_{\mu} ) T^{\mu\nu}_{{\rm EM} ;\nu}
   = - j^{\nu} F^{\mu}_{~~\nu},
\end{eqnarray}
where $\rho$, $P$ and $j^{\mu}$ are the energy density, pressure and electric current,
respectively. The r.h.s. of Eq. (\ref{eq:EOMp}) and (\ref{eq:EOMe}) represent the collision
terms. $C^{(C)\mu}_{pe} = - C^{(C)\mu}_{ep}$ is the collision term for the Coulomb scattering
between protons and electrons. This term leads to the diffusion of magnetic field
and can be neglected in the highly conducting medium in early universe.
On the other hand, the collsion terms for the Thomson scattering between protons (electrons)
and photons are expressed as $C^{(T)\mu}_{p(e) \gamma}$. Because the collision term for
the protons can be neglected compared to that for the electrons, difference in velocities
of protons and electrons will be induced which leads to electric current. This electric
current becomes a source for magnetic field.

Now we evaluate the collision term for the Thomson scattering:
$\gamma(p_{i}) + e^{-}(q_{i}) \rightarrow \gamma(p'_{i}) + e^{-}(q'_{i})$,
where the quantities in the parentheses denote the particle momenta.
The collision integral $C^{(T)}_{\gamma e}[f(p_{i})]$ for this scattering is given as [\cite{KT05}],
\begin{eqnarray}
C^{(T)i}_{\gamma e}[f(p_{i})]
&=& \int \frac{d^{3}p}{(2 \pi)^{3}} p^{i} C^{(T)}_{\gamma e}[f(p_{i})] \nonumber \\
&=& \frac{4 \sigma_{T} \rho_{\gamma} n_{e}}{3}
    \left[ ( u_{e}^{i} - u_{\gamma}^{i} ) + \frac{1}{8} u_{ej} \Pi_{\gamma}^{ij} \right],
\label{eq:collision_term}
\end{eqnarray}
where $\sigma_{T}$ is the cross section of the Thomson scattering.
Here moments of the distribution functions are given by
\begin{eqnarray}
&& \int \frac{d^{3}p}{(2\pi)^{3}} p f_{\gamma}(p_{i}) = \rho_{\gamma}, \\
&& \int \frac{d^{3}p}{(2\pi)^{3}} p^{i} f_{\gamma}(p_{i})
   = \frac{4}{3} \rho_{\gamma} u_{\gamma}^{i}, \\
&& \int \frac{d^{3}p}{(2\pi)^{3}} p^{i} f_{e}(p_{i})
   = \rho_{e} u_{e}^{i}, \\
&& \int \frac{d^{3}p}{(2\pi)^{3}} p^{-1} p^{i} p^{j} f_{\gamma}(p_{i})
   = \frac{1}{6} \rho_{\gamma} \Pi_{\gamma}^{ij} + \frac{1}{3} \rho_{\gamma} \delta^{ij},
\end{eqnarray}
where $\Pi_{\gamma}^{ij}$ is photon anisotropic stress. It should be noted that
the collision term (\ref{eq:collision_term}) was obtained non-perturbatively
with respect to the cosmological perturbation.

Altogether, the Euler equations for protons and electrons are written as
\begin{eqnarray}
&& m_{p} n u_{p}^{\mu} u_{p;\mu}^{i} - e n u_{p}^{\mu} F_{\mu}^{~i}
   = 0,
\label{eq:EOM_p2} \\
&& m_{e} n u_{e}^{\mu} u_{e;\mu}^{i} + e n u_{e}^{\mu} F_{\mu}^{~i} \nonumber \\
&& ~~
   = - \frac{4 \sigma_{T} \rho_{\gamma} n}{3}
       \left[ ( u_{e}^{i} - u_{\gamma}^{i} ) + \frac{1}{8} u_{ej} \Pi_{\gamma}^{ij} \right],
\label{eq:EOM_e2}
\end{eqnarray}
where $m_{p}$ is the proton mass, and the pressure of proton and electron fluids are neglected.
We also assumed the charge neutrality: $n = n_{e} \sim n_{p}$.
Subtracting Eq. (\ref{eq:EOM_p2})
multiplied by $m_{e}$ from Eq. (\ref{eq:EOM_e2}) multiplied by $m_{p}$, we obtain
\begin{eqnarray}
&& - \frac{m_{p}m_{e}}{e}
   \left[ n u^{\mu} \left( \frac{j^{i}}{n} \right)_{;\mu}
          + j^{\mu}
            \left( \frac{m_{p} - m_{e}}{m_{p} + m_{e}} \frac{j^{i}}{en} - u^{i} \right)_{;\mu}
   \right]
\nonumber \\
&& + e n ( m_{p} + m_{e} ) u^{\mu} F^{i}_{~\mu}
   - ( m_{p} - m_{e} ) j^{\mu} F^{i}_{~\mu} \nonumber \\
&& = - \frac{4 m_{p} \sigma_{T} \rho_{\gamma} n}{3}
       \left[ ( u_{e}^{i} - u_{\gamma}^{i} ) + \frac{1}{8} u_{ej} \Pi_{\gamma}^{ij} \right],
\label{eq:Ohm1}
\end{eqnarray}
where $u^{\mu}$ and $j^{\mu}$ are the center-of-mass 4-velocity of the proton and
electron fluids and the net electric current, respectively, defined as
\begin{equation}
u^{\mu} \equiv \frac{m_{p}u_{p}^{\mu} + m_{e}u_{e}^{\mu}}{m_{p} + m_{e}}, ~~~
j^{\mu} \equiv e n (u_{p}^{\mu} - u_{e}^{\mu}).
\end{equation}
Noting the Maxwell equations $F^{\mu\nu}_{;\nu} = j^{\mu}$, we see that
the first term in the l.h.s. of Eq. (\ref{eq:Ohm1}) are suppressed, compared
to the second term, by a factor [\cite{Subramanian94}]
\begin{equation}
\frac{c^{2}}{L^{2} \omega_{p}^{2}}
\sim 3 \times 10^{-34}
     \left( \frac{10^{3} {\rm cm}^{-3}}{n} \right)
     \left( \frac{1 {\rm kpc}}{L} \right)^{2},
\end{equation}
where $c$ is the speed of light, $L$ is a characteristic length of the system
and $\omega_{p} = \sqrt{4 \pi n e^{2}/m_{e}}$ is the plasma frequency.
Second order vector perturbations are contained in the covariant derivative of
the electric current and were evaluated in [\cite{Matarrese04}]. They obtained
the current magnetic field of order $10^{-29} {\rm G}$ at 1Mpc scale.
As we will see below, this is much smaller than a value we obtain in this paper,
which justifies the neglection of the first term in the l.h.s. of Eq. (\ref{eq:Ohm1}).

The third term in the l.h.s. of Eq. (\ref{eq:Ohm1}) is the Hall term which can also
be neglected because the Coulomb coupling between protons and electrons is so tight
that $|u^{i}| \gg |u_{p}^{i} - u_{e}^{i}|$. Then we have a generalized Ohm's law:
\begin{equation}
u^{\mu} F^{i}_{~\mu}
= - \frac{4 \sigma_{T} \rho_{\gamma}}{3 e}
      \left[ ( u_{e}^{i} - u_{\gamma}^{i} ) + \frac{1}{8} u_{ej} \Pi_{\gamma}^{ij} \right]
\equiv C^{i}.
\end{equation}

Now we derive the evolution equation for magnetic field. This can be obtained
from the Bianchi identities $F_{[\mu\nu,\lambda]} = 0$, as
\begin{eqnarray}
0
&=& \epsilon^{ijk} u^{\mu} F_{[jk,\mu]} \nonumber \\
&=& u^{\mu} B^{i}_{~\mu} - \frac{1}{u^{0}} \epsilon^{ijk} C_{j} u^{0}_{~,k}
    + \epsilon^{ijk} C_{j,k} \nonumber \\
& & - ( u^{i}_{~,j} B^{j} - u^{j}_{~,j} B^{i} )
    + \frac{u^{0}_{~,j}}{u^{0}} ( B^{j} u^{i} - B^{i} u^{j} ),
\label{eq:Bianchi}
\end{eqnarray}
where $\epsilon^{ijk}$ is the Levi-Civit\`{a} tensor and
$B^{i} \equiv \epsilon^{ijk} F_{jk}/2$ is comoving magnetic field. We can expand
the photon energy density, fluid velosities and photon anisotropic stress as
\begin{eqnarray}
&& \rho_{\gamma} = \st{0}{\rho}_{\gamma} + \st{1}{\rho}_{\gamma} + \cdots, ~~~
   u_{0} = 1 + \st{2}{u}_{0} + \cdots, \nonumber \\
&& u_{i} = \st{1}{u}_{i} + \st{2}{u}_{i} + \cdots, ~~~
   \Pi_{\gamma}^{ij} = \st{1}{\Pi}_{\gamma}^{ij} + \cdots,
\end{eqnarray}
where the superscripts $(i)$ denote the order of expansion. Remembering that $B^{i}$
is a small quantity, we see that most of the terms in Eq. (\ref{eq:Bianchi}),
other than the first and third terms, can be neglected. Thus we obtain
\begin{eqnarray}
\dot{B}^{i}
&\sim& - \epsilon^{ijk} C_{j,k} \nonumber \\
&\sim& \frac{4 \sigma_{T} \st{0}{\rho}_{\gamma}}{3 e} \epsilon^{ijk}
       \Biggl[ \frac{\st{1}{\rho}_{\gamma ,k}}{\st{0}{\rho}_{\gamma}} 
              \left( \st{1}{u}_{ej} - \st{1}{u}_{\gamma j} \right)
              + \left( \st{2}{u}_{ej,k} - \st{2}{u}_{\gamma j,k} \right) \nonumber \\
&&              + \frac{1}{8} \left( \st{1}{u}_{el,k} \st{1}{\Pi}{}^{l}_{\gamma j}
                                   + \st{1}{u}_{el} \st{1}{\Pi}{}^{l}_{\gamma j,k} \right)
       \Biggr],
\label{eq:B_dot}
\end{eqnarray}
where the dot denotes a derivative with respect to the cosmic time, and we used the fact
that there is no vorticity in the linear order: $\epsilon^{ijk} \st{1}{u}_{j,k} = 0$.
The contributions of the first two terms in Eq. (\ref{eq:B_dot}) were first noticed
in [\cite{Gopal04}]. From this expression, we see that magnetic field cannot be generated
in the linear order. Here it should be noted that the velocity of electron fluid
can be approximated to the center-of-mass velocity at this order,
$\st{1}{u}_{e}^{i} \sim \st{1}{u}^{i}$.

\section{Evaluation of generated magnetic field}

\begin{figure}
\resizebox{\hsize}{!}
{\includegraphics[width=1cm,clip]{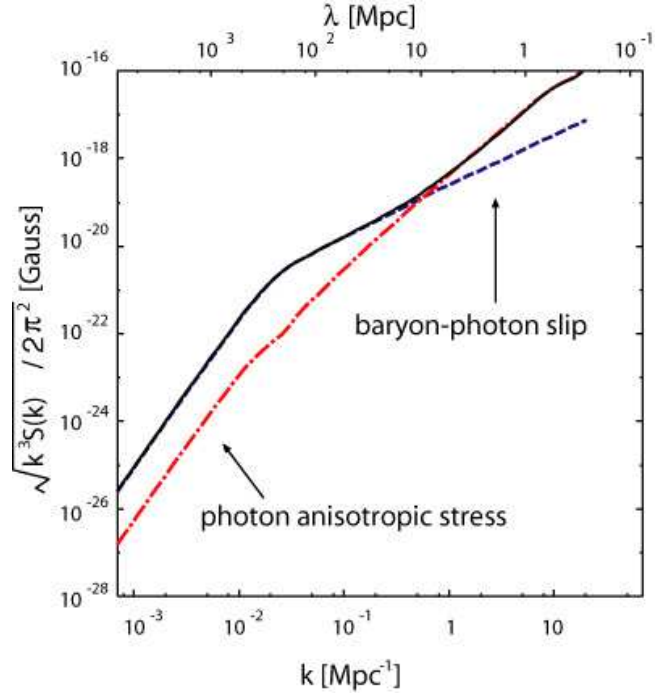}}
\caption{Spectrum of magnetic fields S(k) generated from
cosmological perturbations at cosmological recombination.
(We plot $\sqrt{k^3 S(k)}$ instead of $S(k)$ in order to measure in units of Gauss.) 
Blue dashed and red dot-dashed lines show contributions from the
baryon-photon slip and photon's anisotropic stress, respectively.
The spectrum decays as $k^4$ at scales 
larger than that of the cosmic horizon at cosmological recombination. At
small scales, the contribution from the anisotropic stress of photons
dominates and the spectrum has a slope proportional to $k$.
\label{fig:spectrum}}
\end{figure}

As we saw in Eq. (\ref{eq:B_dot}), there are three main contributions to the generation
of magnetic fields (1) baryon-photon slip term, (2) vorticity difference term, and
(3) anisotropic pressure term.
These terms derive from the fact that electrons are pushed by photons
through Compton scattering when there exist velocity differences between
them, or anisotropic pressure from photons.
Here we now derive the power spectrum of magnetic fields, 
and then perform a numerical calculation to evaluate it.
The power spectrum of magnetic fields $S(k)$ is defined as
$\left<\left|{\vec B}(\vec{k})\right|^2 \right> \equiv S(k)$,
where $\vec{k}$ is the wave
vector, ${\vec B}(\vec{k})$ is Fourier component of magnetic
fields. $S(k)$ represents the expected variance of magnetic
fields ${\vec B}(\vec{k})$, from which the component of the field with
characteristic scale $\lambda$ can be derived through $B_\lambda \approx
\sqrt{k^3 S(k)/(2\pi^2)}$ with $\lambda = 2\pi /k$.
The power spectrum of magnetic fields $S(k)$ is defined by the expected
variance of the Fourier component of magnetic fields ${\vec B}(\vec{k})$ as
$S(k) \equiv \left<\left|{\vec B}(\vec{k})\right|^2 \right>$, where
$\vec{k}$ is the wave vector. The component of the field with
characteristic scale $\lambda$ can then be derived through $B_\lambda \approx
\sqrt{k^3 S(k)/(2\pi^2)}$ with $\lambda = 2\pi /k$.
We consider a standard cosmological model which
consists of photons, baryons, cold dark matter, neutrinos, and
cosmological constant, fixing all the cosmological parameters to the
standard values. 
The density perturbations of them were solved numerically in a range of
scales from $10$ kpc up to $10$ Gpc, and they were then integrated to
obtain $S(k)$.
We found that the field strength of generated magnetic fields at
cosmological recombination can be as large as $10^{-16.8}$ G at 1Mpc
comoving scale, and it becomes even larger at smaller scales
($10^{-12.8}$ G at 10 kpc) (Fig. \ref{fig:spectrum}). 
After cosmological recombination, no magnetic fields would be generated,
since most of electrons were combined into hydrogen atoms and Compton
scattering was no longer efficient.
This means that the fields have an amplitude of $10^{-22.8}$ G at $1$Mpc
($10^{-18.8}$ G at $10$kpc) at present because magnetic fields decay
adiabatically as the universe expands after their generation. 
It is large enough to seed the galactic magnetic fields required by the
dynamo mechanism, which is typically of the order of $10^{-20 \sim -30}$ G
at around the $10$ kpc scale \cite{Widrow02}.

Generated magnetic fields monotonically increase towards smaller scales
on a range of scales in our calculation.
We found that the field has the spectrum $S(k)\propto k^{4}$ at scales
larger than $\sim 10^{2.5}$ Mpc, which corresponds to super-horizon scales at 
recombination, $S(k)\propto k^{0}$ at intermediate scales ($10^{2.5}$
Mpc $<\lambda <10^{1.5}$ Mpc), and  $S(k)\propto k^{1}$ at
scales smaller than $\sim 10$ Mpc, where the contribution from the
anisotropic stress of photons 
dominates. This means that the field strength $B$ is proportional to
$k^2$ at the scales smaller than $1$Mpc.
If the primordial power spectrum of density fluctuations is given
by a simple power law as predicted by inflation,
our result implies that magnetic  
fields with the strength $B \approx 10^{-12.8}$ G unavoidably arise on
$100$ pc comoving scale  at $z \approx 10$. 
This value is interesting for the evolution of structures in the
high-redshift universe, since those magnetic fields would be strong
enough for magneto-rotational instability in the accretion
disks surrounding very first stars (Population III stars) to be
triggered, and affect the transport of their angular
momentum [\cite{MakiSusa04}].
The transport of angular momentum plays an important role for mass
accretion process onto protostars. Since typical mass scale of the Population
III stars is a key [\cite{MakiSusa04}] for the early reionization and
chemical evolution of the universe, cosmologically generated magnetic
fields should be one of essential ingredients in the model of the
structure formation in the high-redshift universe. 

Since magnetic field creation mainly occurs when the modes of density
perturbations with the corresponding scale enter the cosmic horizon
and become causally connected [\cite{KT06}], the magnetic fields should
exist at small scales below $\sim 10$ Mpc even where the Silk damping
effect by photons' diffusion has swept away the density perturbations
at the last scattering epoch. Thus, in 
principle, the detection of magnetic fields below $\sim 10$ Mpc
scales calculated here would tell us about density
perturbations in photons (and baryons) in the early universe even at
scales smaller than 
the diffusion scale at recombination. In this sense, the magnetic field
generated in this mechanism can be regarded as a fossil of density
perturbations in the early universe, whose signature in photons and
baryons has been lost. This result provides consequently a
possibility of probing observations on how density perturbations in
photons had evolved and been swept away at these small scales where no
one can, in principle, probe directly through photons.

\acknowledgements
K. T. and K. I. are supported by a Grant-in-Aid for JSPS Fellows. 
N. S. is supported by a Grant-in-Aid for Scientific Research from the Japanese Ministry
of Education (No. 17540276). This work was supported in part by the Kavli
Institute for Cosmological Physics throught the grant NSF PHY-0114422.

\end{document}